\documentclass[twoside]{photon2007}

\usepackage[latin1]{inputenc}
\usepackage[dvips]{graphicx,epsfig,color}
\usepackage{wrapfig,rotating}
\usepackage{amssymb,amsmath,array}

\newcommand{\sqrtsnn}{\sqrt{s_{_{NN}}}}

\providecommand{\jpsi}{J/\psi}

\providecommand{\ups}{\Upsilon}

\providecommand{\qqbar}{Q\overline{Q}}

\providecommand{\gaga}{\gamma\,\gamma}

\providecommand{\gA}{\gamma\,A}

\providecommand{\elel}{e^+e^-}
\providecommand{\mumu}{\mu^+\mu^-}
\providecommand{\lele}{l^{+}\,l^{-}}

\begin{document}
\title{Quarkonia Photoproduction at Nucleus Colliders}
\author{David d'Enterria$^1$ 
\vspace{.3cm}\\
CERN, PH-EP\\
CH-1211 Geneva 23, Switzerland\\
}

\maketitle

\begin{abstract}
Exclusive photoproduction of heavy quarkonia in high-energy ultraperipheral ion-ion 
interactions ($\gA\rightarrow V\,A$, where $V=\jpsi,\ups$ and the nucleus $A$ 
remains intact) offers a useful means to constrain the small-$x$ nuclear gluon density.
We discuss~\cite{url} preliminary results on $\jpsi$ photoproduction in Au-Au 
collisions at RHIC~\cite{dde_qm05}, as well as full simulation-reconstruction studies of 
photo-produced $\ups$ in Pb-Pb interactions at the LHC~\cite{cms_hi_ptdr}.
\end{abstract}

\section{Introduction}

\begin{figure}[htpb]
\centerline{\includegraphics[width=1.0\columnwidth]{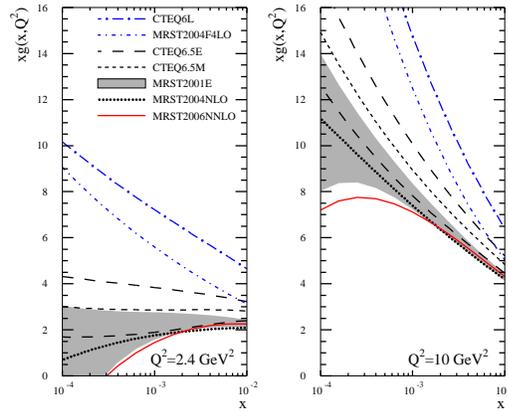}}
\caption{Comparison of recent parametrizations of the low-$x$ gluon distribution  
at scales $Q^2$~=~2.4 (left) and 10~GeV$^2$ (right)~\cite{teubner07}.}
\label{Fig:Gluon}
\end{figure}

\begin{figure}[htpb]
\centerline{\includegraphics[width=0.95\columnwidth]{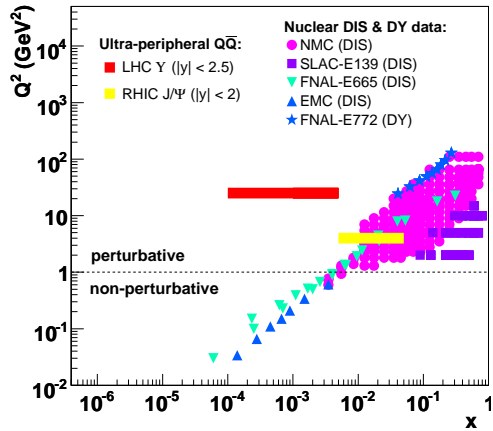}}
\caption{Kinematic ($x,Q^2$) plane probed in e-,$\gamma$-A collisions: 
DIS data compared to ultraperipheral $\qqbar$ photoproduction ranges.}
\label{Fig:Nuclear_Q2x}
\end{figure}

The gluon density, $xG(x,Q^2)$, at small fractional momenta 
$x=p_{\mbox{\tiny{\it parton}}}/p_{\mbox{\tiny{\it proton}}} \lesssim 0.01$
and low, yet perturbative, $Q^2$ is a subject of intensive experimental and theoretical activity. 
On the one hand, DGLAP analyses based on DIS $e$-p data cannot reliably determine $xG$ 
(Fig.~\ref{Fig:Gluon})~\cite{teubner07} as it is only indirectly constrained by the $\log(Q^2)$ 
dependence of the {\em quark} distributions ($F_2$ scaling violations). On the other, there are 
well funded theoretical arguments~\cite{lowx} that support the inapplicability of linear QCD 
(DGLAP- or BFKL-type) evolution equations at low enough values of $x$, due to the increasing 
importance of gluon-gluon fusion processes (``parton saturation''). This regime is theoretically
described e.g. in the Colour-Glass-Condensate~\cite{cgc} or ``black-disk limit''~\cite{Frankfurt:2005mc}
approaches. Our knowledge of the low-$x$ gluon distribution in the {\em nucleus} is even more scarce. 
Nuclear DIS data only cover the range above $x\approx 10^{-2}$ (Fig.~\ref{Fig:Nuclear_Q2x}), 
and gluon saturation effects are expected to be much larger in nuclei than in the proton
due to their larger transverse parton density.

\section{$\qqbar$ photoproduction in ultra-peripheral A-A interactions}

\begin{figure}[htpb]
\centerline{\includegraphics[width=0.9\columnwidth,height=4.cm]{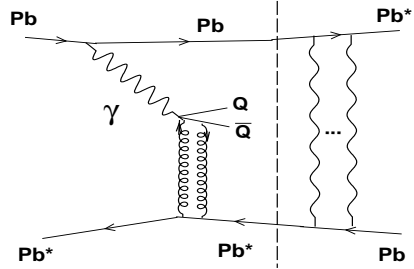}}
\caption{Schematic diagram for diffractive quarkonia photoproduction in $\gA$ collisions.}
\label{Fig:UPCdiag}
\end{figure}

Exclusive quarkonia photoproduction offers an attractive opportunity to constrain 
the low-$x$ gluon density at moderate virtualities, since in such processes the gluon 
couples {\em directly} to the $c$ or $b$ quarks (see Fig.~\ref{Fig:UPCdiag}) and the 
cross section is proportional to the gluon density {\em squared} (see~\cite{teubner07} 
and refs. therein). The mass of the $Q\bar Q$ vector meson introduces a relatively large 
scale, amenable to a perturbative QCD (pQCD) treatment.

High-energy nuclear photoproduction studies are possible in Ultra-Peripheral Collisions
(UPCs) of heavy-ions~\cite{UPCreport} in which the strong electromagnetic fields involved 
are equivalent to the exchange of quasi-real photons 
with maximum energies $\omega_{max}\approx$~3~GeV (100~GeV) at RHIC (LHC).
Correspondingly, the maximum photon-nucleus c.m. energies are of the order 
$W^{max}_{\gA}\approx$~35~GeV (1~TeV) at RHIC (LHC). Thus, in 
$\gA\rightarrow \jpsi \,(\ups)\, A^{(*)}$ processes, the gluon distribution can be probed 
at values as low as $x=M_V^2/W_{\gA}^2\approx 10^{-2} (10^{-4})$ 
(Fig.~\ref{Fig:Nuclear_Q2x}).
Gluon saturation effects are expected to reveal themselves through strong 
suppression of hard exclusive diffraction relative to the leading-twist approximation~\cite{Frankfurt:2005mc}. 
While this suppression may be beyond the kinematics achievable for $J/\psi$ 
photoproduction in UPCs at RHIC, $x\approx 0.01$ and 
$Q^2_{\rm eff} \approx M_V^2/4\approx 3$~GeV$^2$, 
it could be important in UPCs at the LHC~\cite{UPCreport}.

\section{$\jpsi$ photoproduction in Au-Au at RHIC (PHENIX)}

The PHENIX experiment has measured $\jpsi$ photoproduction at mid-rapidity 
in Au-Au UPCs at $\sqrtsnn$~=~200~GeV in the dielectron channel~\cite{dde_qm05}. 
The UPC events were triggered requiring (i) a cluster in the electromagnetic calorimeter 
(EMCal) above 0.8 GeV, (ii) a rapidity gap in one or both $3.0<|\eta|<3.9$ ranges, and
(iii) at least 30~GeV energy deposited in one or both of the Zero-Degree-Calorimeters (ZDCs). 
This last condition very efficiently selects ultra-peripheral events accompanied by forward 
neutron emission ($Xn$) coming from the electromagnetic dissociation of one (or both) 
Au$^\star$ nuclei, which occurs with a large probability, $P_{Xn}\sim$~0.64 (at $y$~=~0) 
at RHIC energies~\cite{starlight_Xn}. 
Electron reconstruction is done combining the central 
tracking devices, Ring Imaging \v{C}erenkov (RICH) counters, and the EMCal. 

\begin{figure}[htpb]
\centerline{\includegraphics[width=0.95\columnwidth]{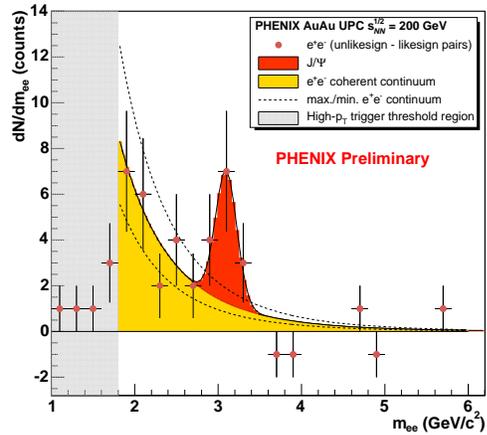}}
\caption{Invariant mass distribution of $e^+e^-$ pairs measured in UPC Au-Au
fitted to the combination of a $\gaga\rightarrow\elel$ continuum 
plus a $\gA\rightarrow \jpsi\,A$  signal~\cite{dde_qm05}.}
\label{Fig:Minv_ee_jpsi}
\end{figure}

\begin{figure}[htpb]
\centerline{\includegraphics[width=0.99\columnwidth,height=5.75cm]{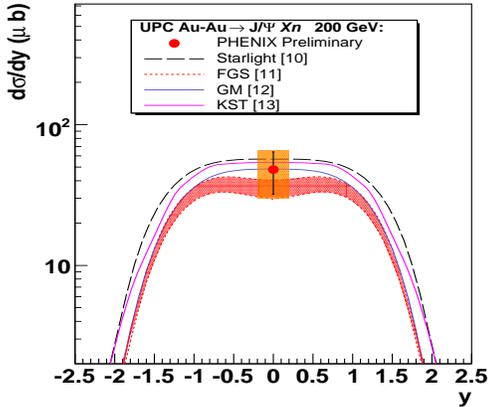}}
\caption{Preliminary cross-section of coherent $\jpsi$ production 
at $y$~=~0 in UPC Au-Au at $\sqrtsnn$~=~200~GeV compared 
to various theoretical calculations~\protect\cite{starlight,strikman05,machado07,kopeliovich07}.}
\label{Fig:dNdy_vs_model}
\end{figure}

The invariant mass distribution of all reconstructed $e^\pm$ pairs is shown in Fig.~\ref{Fig:Minv_ee_jpsi}. 
The plot shows the expected $\gaga\rightarrow\elel$ continuum curve combined with a fit 
to a Gaussian at the $\jpsi$ peak. The total number of $\jpsi$'s is 
$10 \pm 3 \mbox{ (stat) } \pm 3 \mbox{ (syst.)}$, where the systematic 
uncertainty is dominated by the di-electron continuum subtraction. Within the (still large) 
experimental errors, the preliminary $\jpsi$ cross-section of
$d\sigma/dy|_{|y|<0.5}\,=\,48 \pm 14 \mbox{(stat)}\,\pm 16 \mbox{(syst) }\,\mu b$
is consistent with various theoretical predictions~\cite{starlight,strikman05,machado07,kopeliovich07} 
(see Fig.~\ref{Fig:dNdy_vs_model}, where the FGS and KST rapidity distributions have been scaled 
down according to~\cite{starlight_Xn} to account for the reduction of the yield expected when requiring 
coincident forward neutron emission). The band covered by the FGS predictions includes the 
$\jpsi$ cross sections with and without gluon shadowing~\cite{strikman05}. The current experimental 
uncertainties preclude yet any detailed conclusion regarding the nuclear gluon distribution. The possible 
contribution of an additional incoherent ($\gamma$-nucleon$\rightarrow \jpsi$) component -- amounting 
to about $\sim$50\% of the coherent ($\gA$) yield at $y$~=~0~\cite{strikman05} -- should be taken 
under consideration too.

\section{$\ups$ photoproduction in Pb-Pb at the LHC (CMS)}

At the LHC energies, the cross section for $\ups(1S)$ photoproduction in UPC Pb-Pb 
at $\sqrtsnn$~=~5.5 TeV is of the order of  150 $\mu$b~\cite{starlight,Frankfurt:2003qy}. 
Inclusion of leading-twist shadowing effects in the nuclear PDFs reduces the yield by 
up to a factor of two, $\sigma_{\ups}$~=~78~$\mu$b~\cite{Frankfurt:2003qy}. 
Even larger reductions are expected in calculations including gluon-saturation 
(Colour Glass Condensate) effects~\cite{Goncalves:2006ed}.

\begin{figure}[htpb]
\centerline{\includegraphics[width=0.95\columnwidth]{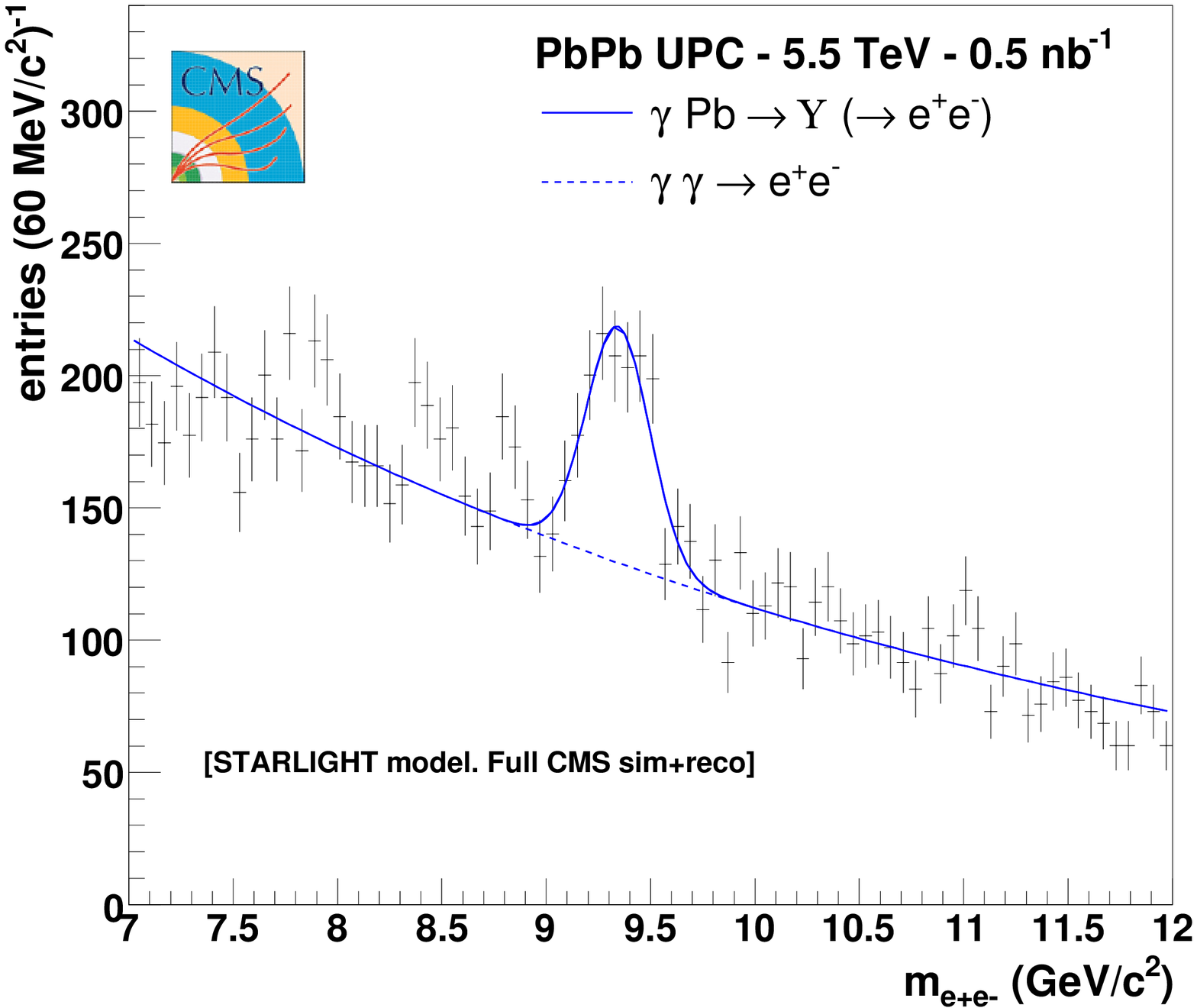}}
\centerline{\includegraphics[width=0.95\columnwidth]{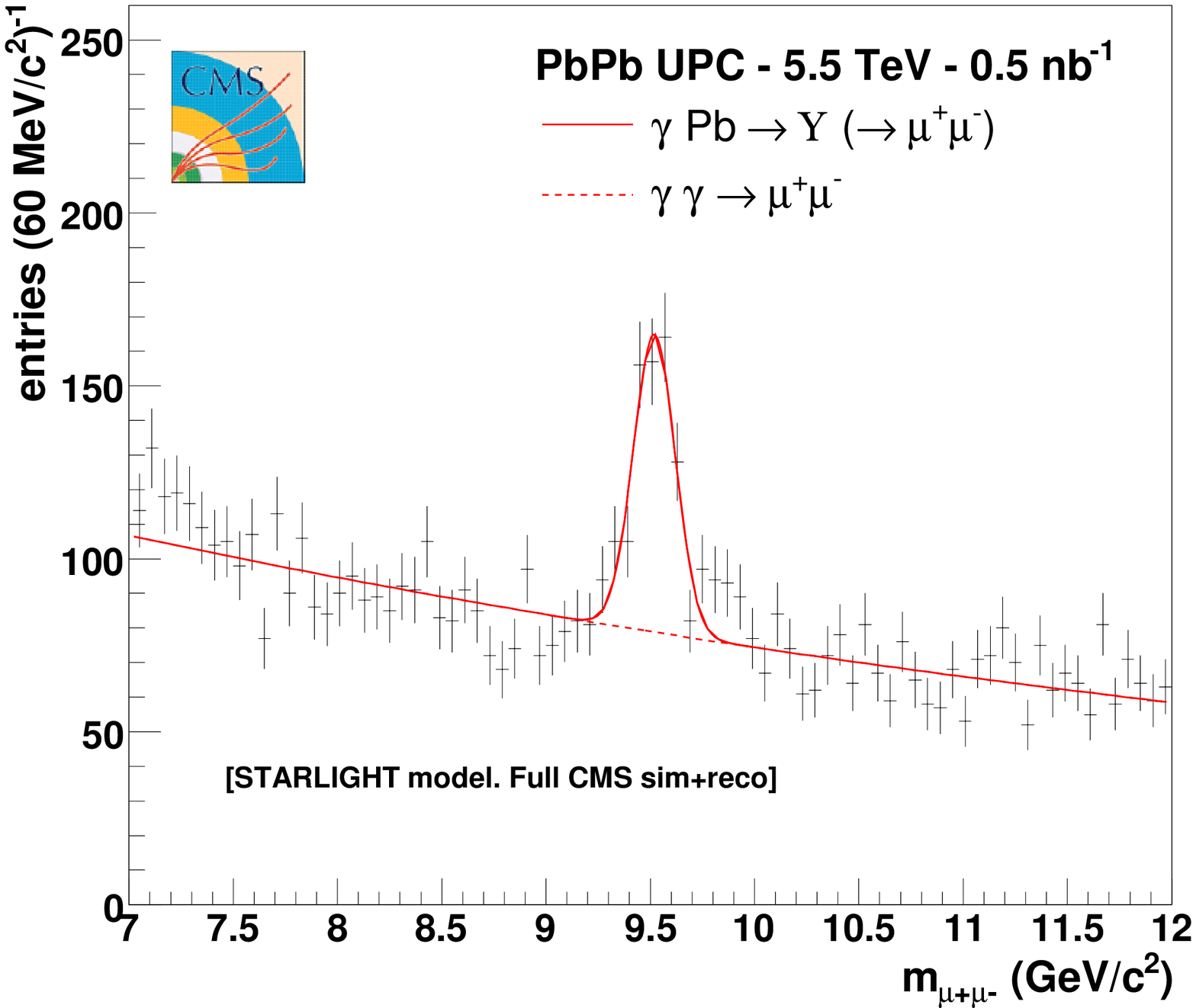}}
\caption{Expected $\elel$ (top) and $\mumu$ (bottom) invariant mass distributions from 
$\gamma\,$Pb$\rightarrow \Upsilon\,$Pb$^\star$ ($\ups\rightarrow \lele$, signal)
and $\gaga\rightarrow \lele$ (background) in UPC Pb-Pb at $\sqrtsnn$~=~5.5 TeV in CMS.}
\label{Fig:upc_ups_cms}
\end{figure}

Full simulation+reconstruction studies~\cite{cms_hi_ptdr} of input distributions generated with the 
{\sc starlight} MC~\cite{starlight} have shown that CMS can measure $\ups\rightarrow \elel$, $\mumu$ 
within $|\eta|<$~2.5, in UPCs tagged with neutrons detected in the ZDCs~\cite{zdc_cms}, 
with large efficiencies ($\epsilon_{\rm rec}\times {\cal A}$cc$\times \epsilon_{\rm yield-extract}\approx$~20\%).
Figure~\ref{Fig:upc_ups_cms} shows the reconstructed $dN/dm_{\lele}$ around the $\ups$ mass 
(only the ground-state, $\ups(1S)$, of the bottomonium family was generated). The signal over 
continuum background is around one for both decay modes. 
The total expected number of $\ups$ events, normalised to the nominal 0.5 nb$^{-1}$ Pb-Pb integrated luminosity,
is $\sim$\,500, and the $p_T$ resolution is good enough to separate the coherent (peaked at 
very low $p_{\rm T}\approx M_{V}/\gamma\approx$~30~MeV/$c$) from the incoherent 
components. With such a statistics, detailed $p_T$,$\eta$ studies 
can be carried out, that will help constrain the low-$x$ gluon density in the nucleus.

\section{Summary}

High-energy quarkonia photoproduction provides a particularly useful means to
constrain the poorly known low-$x$ gluon distribution of the nucleus in the clean 
environment of ultra-peripheral (electromagnetic) ion-ion collisions. 
Gluon saturation effects in the small-$x$ domain of the nuclear 
wavefunction are expected to result in a suppression of hard exclusive 
diffraction yields relative to linear QCD expectations. We have presented preliminary 
PHENIX results of exclusive $\jpsi$ photoproduction in 200-GeV Au-Au interactions,
as well as the perspectives of the CMS experiment in 5.5-TeV Pb-Pb collisions 
at the LHC. In the absence of strong non-linear QCD effects, 
around 500 photo-produced $\ups$ will be reconstructed in the CMS acceptance 
with nominal integrated luminosities.

\section*{Acknowledgments}

My gratitude to the organisers of {\sc photon}'07 conference -- in particular Gerhard Baur 
and Maarten Boonekamp -- for their kind invitation and for the stimulating programme. 
Special thanks due to Joakim Nystrand for providing the {\sc starlight} 
predictions and for useful discussions, as well as to Mark Strikman for informative exchanges. 
Work supported by the 6th EU Framework Programme contract MEIF-CT-2005-025073.


\section{Bibliography}

\def\IJMPA{{Int. J. Mod. Phys.}~{\bf A}}
\def\EPJ{{Eur. Phys. J.}~{\bf C}}
\def\JPG{{J. Phys.}~{\bf G}}
\def\JHEP{{J. High Energy Phys.}~}
\def\NCA{Nuovo Cimento~}
\def\NIM{Nucl. Instrum. Methods~}
\def\NIMA{{Nucl. Instrum. Methods}~{\bf A}}
\def\NPA{{Nucl. Phys.}~{\bf A}}
\def\NPB{{Nucl. Phys.}~{\bf B}}
\def\PLB{{Phys. Lett.}~{\bf B}}
\def\PLC{Phys. Repts.\ }
\def\PRL{Phys. Rev. Lett.\ }
\def\PRD{{Phys. Rev.}~{\bf D}}
\def\PRC{{Phys. Rev.}~{\bf C}}
\def\ZPC{{Z. Phys.}~{\bf C}}

\begin{footnotesize}

\end{footnotesize}


\end{document}